\documentclass[twocolumn]{aa} 

\usepackage{graphicx}
\usepackage{txfonts}
\usepackage{amssymb}
\usepackage{natbib}
\usepackage{xspace}
\newcommand{\un}[1]{\,\ensuremath{\mathrm{#1}}}
\newcommand{\am}{$^{\prime}$\xspace}

\newcommand{\integ}{{\it INTEGRAL}\xspace}
\newcommand{\ibis}{IBIS\xspace}

\newcommand{\isgri}{ISGRI\xspace}

\newcommand{\xmm}{{\it XMM-Newton}\xspace}
\newcommand{\sax}{{\it BeppoSAX}\xspace}
\newcommand{\rxte}{{\it RXTE}\xspace}
\newcommand{\asca}{{\it ASCA}\xspace}
\newcommand{\rosat}{{\it ROSAT}\xspace}

\def\etal{{et~al.\ }}
\def\ie{{\em i.e.\ }}
\newcommand{\gammaray}{$\gamma$-ray\xspace}

\newcommand{\xray}{X-ray\xspace}
\newcommand{\xrays}{X-rays\xspace}
\newcommand{\coma}{Coma~Cluster\xspace}
\newcommand{\icm}{ICM\xspace}

\newcommand{\hxr}{HXR\xspace}

\begin{document}

\title{Coma revealed as an extended hard \xrays source by INTEGRAL IBIS/ISGRI}
\author{M. Renaud\inst{1,2}, G. B\'elanger\inst{1,2}, J. Paul\inst{1,2}, F. Lebrun\inst{1,2}, R. Terrier\inst{2}}

\offprints{M. Renaud : mrenaud@cea.fr}

\institute{Service d'Astrophysique, CEA-Saclay, 91191, Gif-Sur-Yvette, France 
\and APC-UMR 7164, 11 place M. Berthelot, 75231 Paris, France}

\date{Received 8 March 2006 / Accepted 27 April 2006}
\authorrunning{M. Renaud et al.}
\titlerunning{\integ/\ibis observations of the \coma}

\abstract{}{We report the \integ/\ibis observations of the \coma in the hard 
\xray/soft \gammaray domain.}{Since the \coma appears as an extended source, 
its global intensity and significance cannot be directly extracted with standard 
coded mask analysis. We used the method of imaging the extended sources with 
a coded mask telescope developed by Renaud \etal (2006).}{The imaging capabilities 
and the sensitivity of the \ibis/\isgri coded mask instrument allows us to identify 
for the first time the site of the emission above $\sim$ 15\un{keV}. We have studied 
the \coma morphology in the 18-30\un{keV} band and found that it follows the 
prediction based on \xray observations. We also bring constraints on the non-thermal 
mechanism contribution at higher energies.}
{}


 \maketitle


\section{Introduction}
\label{s:intro}

Clusters of galaxies are the largest bound structures in the visible universe, 
and amongst the most luminous ones. This makes them important cosmological probes. 
The intercluster medium (\icm) is heated to high temperatures by the initial collapse.
In addition, it is believed that as groups and smaller clusters of galaxies merge to form 
larger ones, violent shocks compress and heat the intercluster medium to \xray 
emitting temperatures ($10^7$--$10^8$\un{K}). These shocks could also drive the 
acceleration of ions and electrons to relativistic energies, thereby producing a spectrum 
extending from the radio to the \gammaray domains through processes such as synchrotron 
emission and inverse Compton (hereafter, IC) scattering.

The observations of diffuse radio halos from cluster cores \cite{c:giovannini93}, and the 
more recent reported detection of extreme ultraviolet (EUV) and hard \xray (\hxr) 
emission from clusters \citep{c:bowyer99,c:fusco99,c:rephaeli99} seem to point to the 
possible presence of non-thermal and/or supra-thermal particles. Although the origin 
of the radio halos is most probably synchrotron emission from high energy electrons, that of 
the EUV photons and hard \xrays has been the source of much debate and is still uncertain. 
Sarazin (1999) suggested that the extreme UV and hard \xrays could arise from IC scattering 
of the energetic, synchrotron producing electrons off the cosmic microwave background 
radiation (see also Bykov \etal 2000 for a detailed discussion of nonthermal high energy 
emission in galaxy clusters). Alternatively, this emission could also be due to bremsstrahlung 
radiation of supra-thermal electrons accelerated by turbulent gas dynamics in the \icm 
\cite{c:ensslin99,c:sarazin00}. If the \hxr cluster component is confirmed, then the question 
remains: is it thermal or non-thermal in origin?

Thus far, \xray imaging observations of clusters such as Coma, have shown that the 
emission in the \xray range is predominantly thermal, originating from bremsstrahlung 
in the hot \icm. \sax and \rxte detected a hard \xray excess that extends beyond 
10\un{keV} and apparently deviates from the steep thermal spectrum expected from 
pure bremsstrahlung emission \cite{c:fusco99,c:fusco04,c:rephaeli99,c:rephaeli02}. 
Given that these observations were carried out with non-imaging instruments, 
contamination from point sources contained in their field of view is not unlikely. 
Moreover, a recent second look at the \sax data on Coma seems to show that there is 
no evidence of a \hxr excess \cite{c:rossetti04}. This debate can only be convincingly 
resolved and put to rest through sensitive imaging observations at high energies 
beyond the thermally dominated \xray emission.

In this letter, we report on hard \xray/soft \gammaray observations of the \coma with 
the \ibis/\isgri \cite{c:ubertini03,c:lebrun03} coded mask instrument onboard \integ 
\cite{c:winkler03}. A large field of view (29$^{\circ}$\,$\times$\,29$^{\circ}$, 
8$^{\circ}$\,$\times$\,8$^{\circ}$ fully coded), fine imaging capabilities 
(PSF of 12\am FWHM, Gros \etal 2003), and an unprecedented 3$\sigma$ broadband 
sensitivity of $\sim$\,1\un{mcrab} at 20\un{keV} ($10^5$\un{s}, $\Delta$E\,=\,E/2), 
make \ibis/\isgri ideal for probing the nature of the \hxr excess in the \coma: {\bf the 
first extended source detected with  this instrument}. 


\section{Observations and Data Reduction}
\label{s:obs}

The \coma was observed with \integ in 2003 during revolutions 36, 71 and 72. Almost 250 
pointings aiming within 13$^{\circ}$ of Coma amount to $\sim$500\un{ks}. Using these data, 
Krivonos \etal (2005) demonstrated the ability of \ibis/\isgri to detect faint extra-galactic sources, 
and established a catalog of 12 serendipitous sources ($\geq$\,4$\sigma$) in addition to the \coma. 
The authors noted that the source corresponding to the latter was extended, and hence, 
its flux and significance calculated with the standard analysis were likely inaccurate. 
Renaud \etal (2006) presented a general method to determine the flux and the total detection 
significance for any extended celestial source observed through a coded mask. We here 
apply this method to the study of the spatial and spectral features of the \coma.

Reconstructed sky images with coded mask telescopes are correlation maps between the 
detector image and a decoding array derived from the mask pattern. The deconvolution algorithm 
implemented in the \integ Off-Line Scientific Analysis (OSA, Goldwurm \etal 2003) is optimized 
for point-sources and the flux of any source is given by the peak of the SPSF at its position in 
the correlation map. For this reason, the intensity of an extended source cannot be derived directly 
using the standard processing. This limitation can be overcome by constructing images of flux per 
solid angle (\ie per sky pixel), in which the global flux is given by the sum of intensities over
the emissive region. The details of this method are presented in Renaud \etal (2006).

We analyzed the \coma data with OSA v.\,5.0 and obtained individual sky images in the 
18--30, 30--50, 50--100 and 100--150\un{keV} energy bands, directly from the standard analysis. 
We evaluated their quality by measuring the noise given by the width of the distribution of 
significance values, and then constructed images in flux per sky pixel which were combined to 
make mosaics in the four energy ranges.


\section{Results}
\label{s:res}

\subsection{Morphology of the emission}
\label{s:geom}

\begin{figure}[htb]
\centering
\includegraphics[scale=0.42]{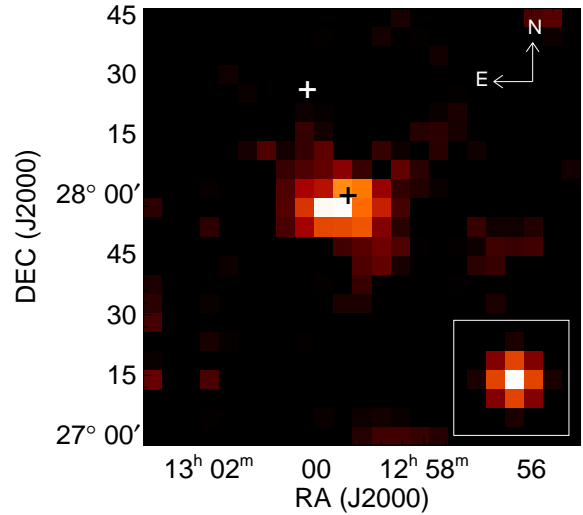}

\caption{\ibis/\isgri significance image of the Coma cluster in the 18--30\un{keV}
energy range. The black cross indicates the position of NGC~4874,
near the center of the \xray emission from Coma \cite{c:briel01}.
The white cross shows the location of X~Comae, a bright Seyfert galaxy
at $\sim$30$^{\prime}$ from the cluster. The colour scale is linear, with black
corresponding to 1$\sigma$ and white to 8$\sigma$. 
The \ibis SPSF is also shown in the lower right corner.}

\label{f:ima_isgri}
\end{figure}

The \ibis/\isgri significance map of the \coma in the energy range 18--30\un{keV} is 
shown in Figure\,\ref{f:ima_isgri}. We have overlaid the instrument's response to a 
point-like source of comparable intensity in the bottom right corner of the image. This 
serves as a morphological comparison, and strongly suggests that the \coma is indeed 
an extended source for \ibis/\isgri. The maximal pixel value within the extended 
emission is at the $\sim$8$\sigma$ confidence level, consistent with Krivonos \etal 
(2005) who noted that it was likely unresolved.

In the \xray domain, the spectrum of the \coma  is dominated by thermal 
bremsstrahlung from hot intercluster gas. \rxte observations \cite{c:rephaeli02} 
yielded values from which we can derive the expected ratio of non-thermal to 
thermal emission at higher energies. These are 0.2 and 1.0 in the 18--30\un{keV} 
and 30--50\un{keV} energy ranges respectively. \xmm observations revealed marked 
temperature variations (from 4 to 11\un{keV}) at distances $>$10\am from 
the centre of the cluster (Arnaud \etal 2001, Briel \etal 2001, Neumann \etal 2003). 
A hot front in the southwest (kT $\sim$ 11\un{keV}) was interpreted as an adiabatic 
compression due to recent accretion of matter. Within the core, centered on 
NGC~4874, these authors found a homogeneous temperature distribution of
$\sim$8.25\un{keV}. 

\begin{figure}[htb]
\centering
\includegraphics[scale=0.42, angle=-90]{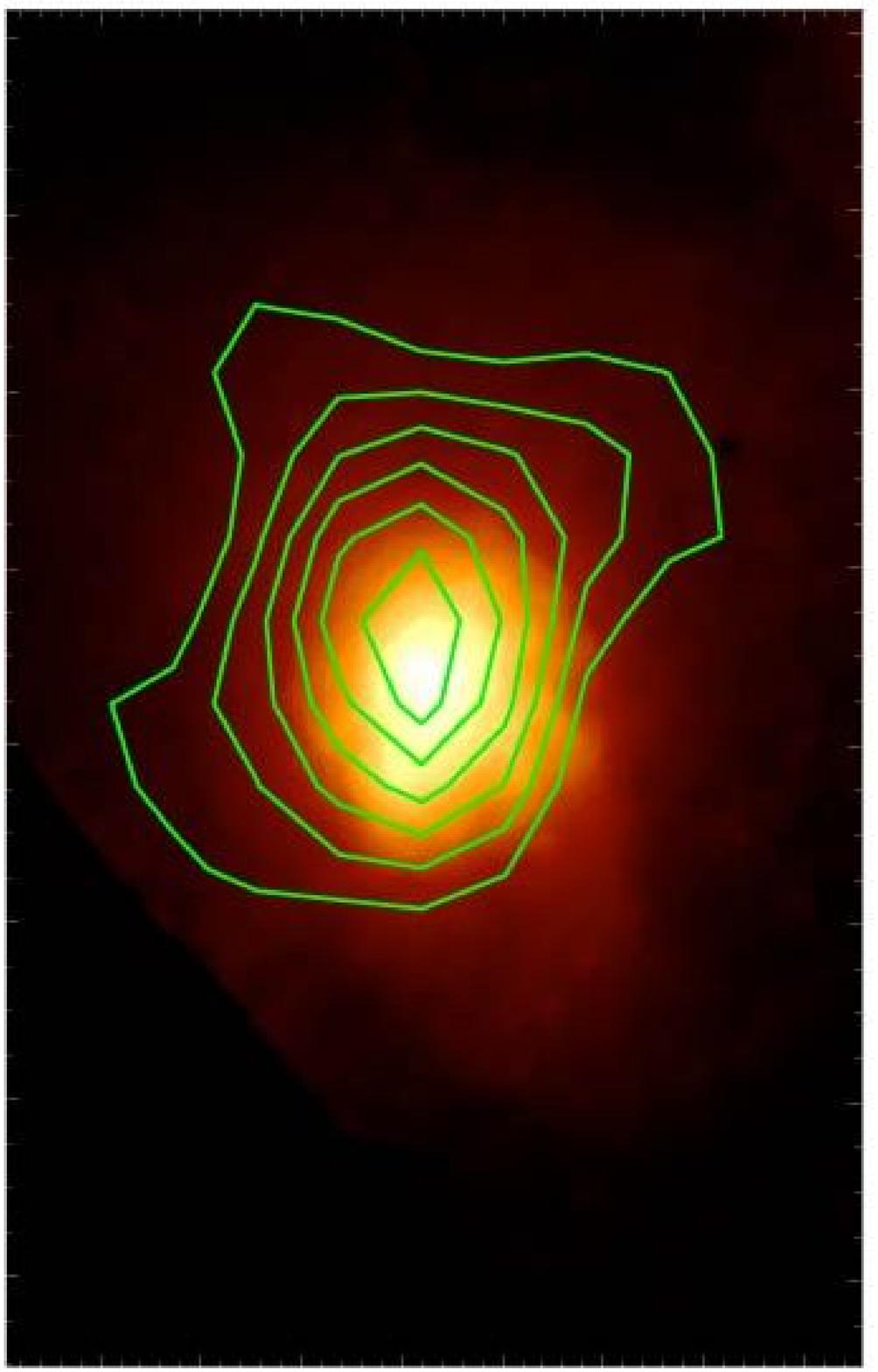}
\includegraphics[scale=0.28]{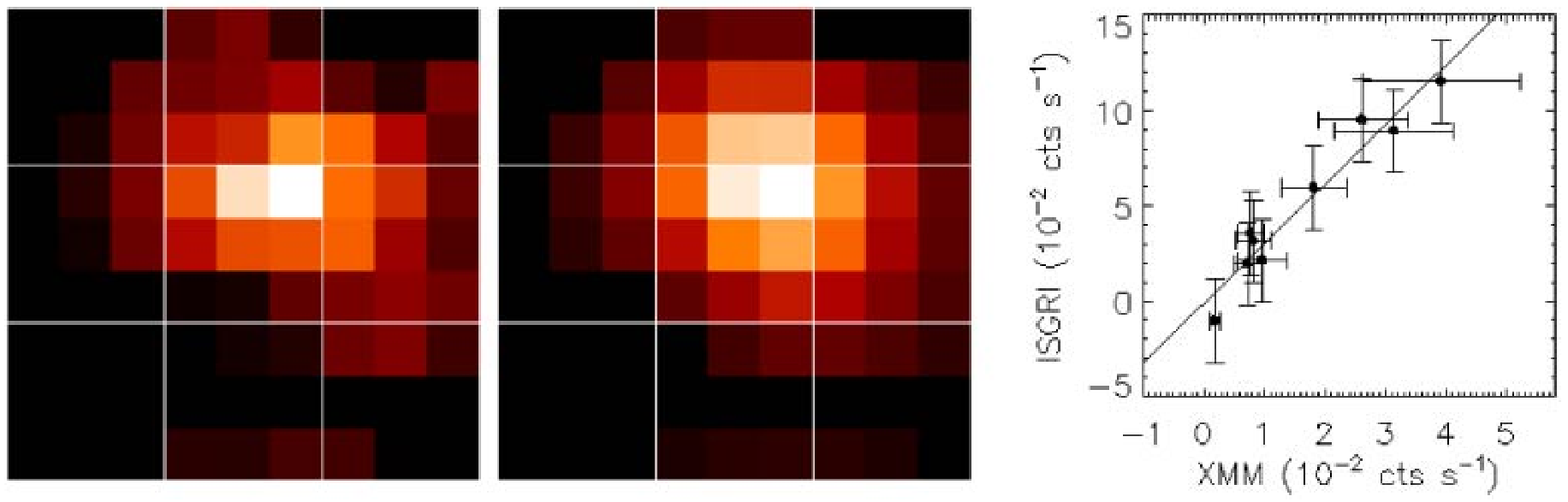}

\caption{{\em Top panel:} \xmm 0.3-2\un{keV} mosaic image \cite{c:neumann03}
with \ibis/\isgri significance contours in the 18--30\un{keV} energy range, 
going from 2$\sigma$ to 10$\sigma$. 
{\em Bottom panel:} \ibis/\isgri image of flux per pixel in the 18--30\un{keV} 
band centered on the \coma (left), the predicted morphology based on the \xmm 
images in intensity and inferred temperature (middle), and the correlation between 
the 18--30\un{keV} and 0.3--2\un{keV} images (right). The white squares delineate
the regions over which the flux was summed for the purpose of correlation study.}

\label{f:images}
\end{figure}

Given that the \xray emission between 0.3 and 2\un{keV} depends only slightly 
on the plasma temperatures in the range of those measured in the ICM 
(Fig.\,1 of Arnaud 2005), the \xmm map in this energy range is a good estimate 
of the ICM gas emissivity. Therefore, we can use the 0.3--2\un{keV} 
EPIC MOS mosaic image, shown in the top panel of Figure~\ref{f:images} and
overlaid with the \ibis/\isgri 18--30\un{keV} contours, in conjunction with the 
temperature map of Arnaud \etal (2001) to construct a map of the expected
thermal emission  in the 18--30\un{keV} band. Moreover, for best
accuracy, we computed correction factors, which account for the temperature 
variations across the cluster, using the {\footnotesize\texttt{mekal}} model in 
XSPEC v.11.3. The redshift and abundances were set to $z$\,=\,0.0231 and 0.25, 
respectively \cite{c:arnaud01}. The relative intensity of each region was weighted 
with the corresponding correction factor. The final 18--30\un{keV} map, convolved 
with the \ibis/\isgri PSF, is shown in the middle bottom panel of Figure~\ref{f:images}. 
The \ibis/\isgri image shown in Figure~\ref{f:ima_isgri} is reproduced in the bottom 
left panel for reference.

To calculate the correlation coefficient between the \ibis/\isgri mosaic in the
18--30\un{keV} range and the expected map for the emission in this energy band
constructed in the manner described above, the sky pixels over which the flux
is summed have to be independent. This is very nearly the case at all scales larger
than the size of the SPSF, which is 12\am or 2.5 sky pixels. The white squares in 
the bottom left and middle panels of Figure~\ref{f:images} correspond to 
regions that are 3 pixels (14.$^{\prime}$5) by 3 pixels in size, and thus regions that
we can consider independent from one another as far as the flux and variance are
concerned. By summing the flux per pixel over the nine regions in each map, 
we obtained two sets of nine intensities. The uncertainties associated with the
\ibis/\isgri map were derived according the prescription of Renaud \etal (2006),
and those of the interpolated map were calculated based on an uncertainty of
1\un{keV} on the plasma temperature (Arnaud 2006, priv. comm.).

We found a linear correlation coefficient of $\rho$\,=\,0.95, and show the 
correlation plot in the bottom right panel of Figure~\ref{f:images}. 
The probability that a random sample of 9 uncorrelated data points yield a 
linear correlation coefficient of 0.95 or larger is 8.7\,$\times$\,10$^{-5}$.
This close agreement between the morphologies of the detected and expected 
thermal emission from the \coma at energies between 18 and 30\un{keV} brings 
us to conclude that what is seen in this energy range by \ibis/\isgri is compatible
with an emission of pure thermal origin with no indication of a non-thermal 
component having a significantly different morphology. A distinction between 
thermal and non-thermal emission components could be detected only through a 
fine spectro-imaging analysis on small spatial scales and over the entire emissive 
region. It is interesting to note that this result adds weight to the temperature
variations seen in \xrays by \xmm.


\subsection{Reconstructed global intensity}
\label{s:flux}

We now turn to the question of the global intensity of this emission when the 
source is considered to be extended. Since instruments such as \sax/PDS and RXTE 
are only capable of recording the total emission within their $\sim$1$^{\circ}$ FOV, 
their detection of a HXR tail from the \coma could not definitively and exclusively 
be attributed to it. This uncertainty is substantiated by the presence of other 
sources present in the FOV like X~Comae, a bright Seyfert galaxy discovered by 
\rosat/PSPC $\sim$30\am northeast of the \coma \cite{c:dow95}. Even if X~Comae may 
experience flux variability, Fusco-Femiano \etal (1999) demonstrated that because of 
its steep spectral index, an unusually strong variability would be necessary in order 
to have a significant effect on the total detected flux. Hence, its hard \xray emission 
would be blended into the global flux and not detected as variable. Nonetheless, 
since \ibis/\isgri does not detect an excess above 3$\sigma$ at the position of X~Comae 
in either of the four energy bands defined in \textsection\,\ref{s:obs}, we assume that 
the spectra obtained by \rxte and \sax are truly representative of the emission from 
Coma and that there really exists a HXR tail, despite the ongoing dispute on this issue
\cite{c:rossetti04}.

\begin{figure}[ht]
\centering
\includegraphics[scale=0.36]{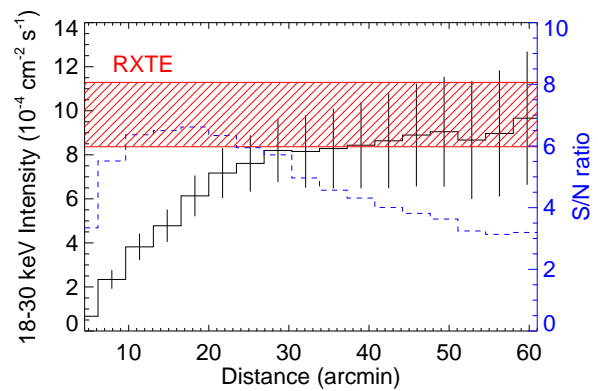}

\caption{The 18--30\un{keV} \ibis/\isgri integrated intensity (left scale, solid line 
with error bars) and corresponding significance calculated on concentric circles 
and expressed as a function of radius (right scale, dashed line). The shaded area 
shows the intensity within the 1$\sigma$ error derived from \rxte observations 
\cite{c:rephaeli02,c:rossetti04}.} 

\label{f:fluxComa}
\end{figure}

Figure \ref{f:fluxComa} shows the intensity in 18--30\un{keV} range, integrated 
over circular areas with increasing radii centred on the pixel of maximum 
intensity located at
RA(J2000) = 12$^{h}$59$^{m}$31.7$^{s}$ and DEC(J2000) = 27$^{\circ}$57\am07.$''$6.
The intensity is shown as the monotonically increasing black curve with error bars.
The dashed blue curve represents the total statistical significance (S/N ratio) calculated
on the basis of each circular area. This analysis is done without any  a priori assumption
on the size of the emitting region. The shaded area shows the \rxte flux taken from 
Rephaeli \& Gruber (2002, Table 1) and Rossetti \& Molendi (2004), whose results are
consistent with one another. The best agreement is found for $R$\,=\,30\am. This value is
also in line with the extent of the \xray surface brightness observed by \asca/GIS 
\cite{c:honda96}. The \ibis imaging sensitivity was deduced from Crab Nebula 
observations in each of the four energy bands assuming a Crab spectrum of 
9.7\,E$^{-2.1}$\un{cm^{-2} s^{-1} keV^{-1}} \cite{c:krivonos05}. 

\begin{figure}[htb]
\centering
\includegraphics[scale=0.28, angle=-90]{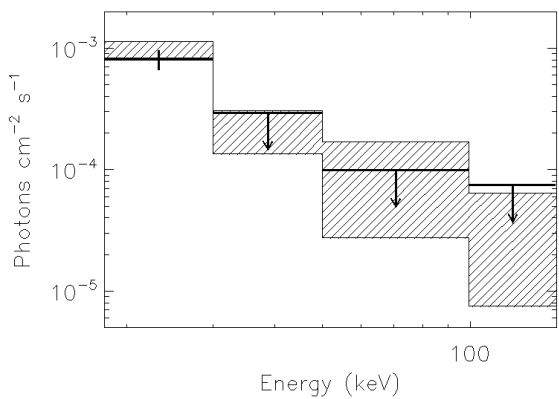}

\caption{\ibis/\isgri spectrum of the \coma. The upper limit in the 30--50\un{keV}
range is calculated assuming that the source has the same shape as it does in the 
18--30\un{keV} energy range. At higher energies, the upper limits are derived as for
a point-source. The best-fit \rxte spectrum within its 1$\sigma$ uncertainties, is 
shown as the dashed area. Upper limits are given at the 3 $\sigma$ confidence level.} 

\label{f:specComa}
\end{figure}

The treatment in the 30--50\un{keV} energy range was performed in the same way
as for the 18--30\un{keV} range, and the flux was summed over the same region
but no significant excess was found. Although we expect that at least half of
the global intensity in the 30--50\un{keV} band arises from thermal processes, the 
emission above 50\un{keV} is purely non-thermal. However, no imaging telescopes have 
observed the Coma at energies $>$10\un{keV} and thus the morphology of the
yet undetected non-thermal emission above 30\un{keV} is unknown.
We derived  upper limits for the flux in the 50--100\un{keV} and 100--150\un{keV}
ranges in the standard way, relying on the assumption that the source is 
point-like ($\Phi$\,$\la$\,8\am). Even though we have shown that this is not so, 
these upper limits are a good estimate in the case of a non-detection.
The 18--150\un{keV} \isgri spectrum is shown in Figure\,\ref{f:specComa}.
For comparison, we have plotted the \rxte best-fit spectrum of this source 
\cite{c:rephaeli02,c:rossetti04} as the shaded area (1$\sigma$ errors), 
and find that the spectra agree.

\section{Conclusion}
\label{s:concl}

The unequalled imaging capabilities and sensitivity of the \ibis/\isgri telescope
on \integ give us the opportunity of imaging extended sources in the hard \xray 
and soft \gammaray domains for  the first time. We have shown that the morphology of 
the emission from the \coma in the 18--30\un{keV} energy range is akin to that in 
\xrays when the temperature variations measured by \xmm across the cluster's emission 
are taken into account. There is no evidence for the presence of non-thermal emission 
arising from a region with the same morphology as that associated with the thermal 
emission. We find that the integrated intensity is in good agreement with previous 
\rxte and \sax observations of Coma.

Future \integ observations will surely reveal more details about the morphology
and nature of the emission from Coma. Our analysis was based on the
assumption that the morphology of the emission is the same in the 30--50\un{keV} range
as it is between 18 and 30\un{keV}. Given that the upper limit we quote for the
30--50\un{keV} flux is only a factor of 1.5 above the mean \rxte spectrum, and that 
non-thermal mechanisms such as IC scattering and non-thermal bremsstrahlung are 
expected to  contribute half the flux in this range, deeper \integ observations
($\sim$1.5\un{Ms}) will permit a  fine morphological study of the emissive region.
This will provide the opportunity to definitively determine whether there is or not 
a \hxr tail, and if so to identify the site of particle  acceleration in the \coma for 
the first time.


\begin{acknowledgements}

We would like to thank M. Arnaud and D. Neumann for providing the \xmm images and 
for helpful suggestions. The present work is based on observations with \integ, an ESA 
project with instruments and science data center (ISDC) funded by ESA members states (especially 
the PI countries: Denmark, France, Germany, Italy, Switzerland, Spain, Czech Republic and Poland, 
and with the participation of Russia and the USA). \isgri has been realized and is maintained in 
flight by CEA-Saclay/DAPNIA with the support of the French Space Agency CNES.

\end{acknowledgements}

\end{document}